\documentclass[]{interact}

\usepackage{epstopdf}
\usepackage[caption=false]{subfig}

\usepackage[numbers,sort&compress]{natbib}
\bibpunct[, ]{[}{]}{,}{n}{,}{,}
\renewcommand\bibfont{\fontsize{10}{12}\selectfont}
\makeatletter
\def\NAT@def@citea{\def\@citea{\NAT@separator}}
\makeatother

\theoremstyle{plain}

\theoremstyle{definition}

\theoremstyle{remark}

\begin{document}

\articletype{ORIGINAL ARTICLE}

\title{Flow Driven Oil Recovery Enhanced with Structural Disjoining Pressure}

\author{
\name{ Shane Laibach\textsuperscript{a}, 
       Egor Vinogradov\textsuperscript{a},
       Jasper Stedman\textsuperscript{a},
       M G Guru Aravindan\textsuperscript{a},
       Myles Geise \textsuperscript{a},
       Viet Sang Doan\textsuperscript{a},
       Sangwoo Shin\textsuperscript{a}, and
       Craig Snoeyink\textsuperscript{a}\thanks{CONTACT C. Snoeyink Email: craigsno@buffalo.edu}}
\affil{\textsuperscript{a} Mechanical and Aerospace Engineering, University at Buffalo, Buffalo, NY USA}
}

\maketitle

\begin{abstract}
Nanofluids have the potential to enhance oil recovery through the structural disjoining pressure, a pressure developed when nanoparticles concentrate at the three-phase contact line. A model microfluidic porous network is used to measure the percentage of oil displaced from this channel as the volume fraction of a Triton X-100 micelle nanofluid is varied from 0 - 30\%. The percentage of oil displaced varies nearly linearly with micellar nanoparticle volume fraction starting with 39\% using deionized water and 89\% using a volume fraction of 30\%. While the trend is clear, significant variability between experiments was observed for a fixed nanofluid volume fraction. This indicates that surface energy heterogeneity is important for the nanofluid oil displacement performance. 
\end{abstract}


\section{Introduction}\label{sec:intro}

\begin{figure}[h]%
\centering
\includegraphics[width=0.7\textwidth]{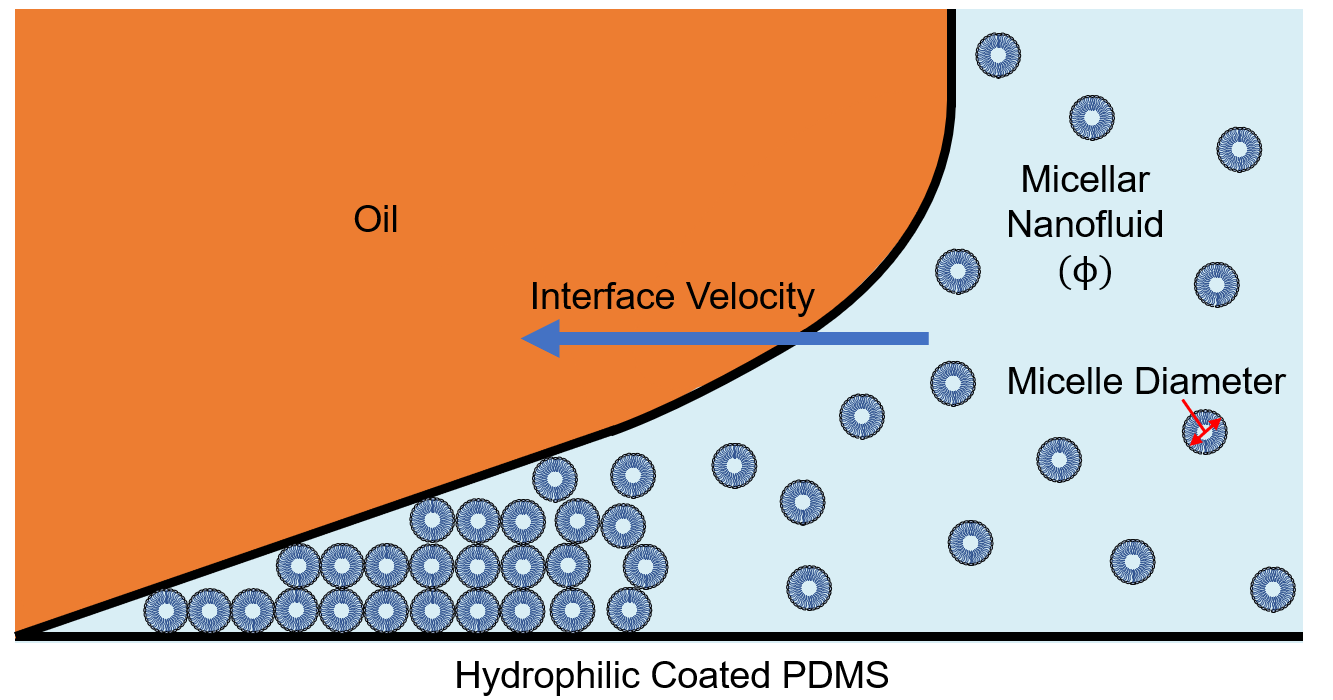}
\caption{Schematic showing increasing concentration of micelles at the three-phase contact line creating an organized nanofluid wedge. This concentration of nanoparticles creates a pressure that drives the interface, displacing the oil.    }
\label{fig:physics}
\end{figure}

Nanofluids have been shown to be a promising method of oil removal in porous media, potentially impacting enhanced oil recovery or oil remediation efforts \cite{zhang_enhanced_2016,wu_cleansing_2013,suleimanov_nanofluid_2011,hendraningrat_stabilizer_2015,hendraningrat_coreflood_2013,contreras-mateus_applications_2022}.  The nanoparticles within these fluids improve oil removal by changing the rheology of the fluid, reducing interfacial tension or creating visco-elastic interfaces, and altering the wettability of the surfaces. These modifications enable the flooding of nanofluid to displace the oil and overcome viscous fingering and large capillary forces which can trap the oil. 

Less studied is the unique ability of concentrated nanoparticles to drive the motion of contact lines through the structural disjoining pressure. Nanoparticles at relatively high volume fractions will concentrate at contact lines as shown in Figure \ref{fig:physics} and produce a pressure, the structural disjoining pressure (SDP), which can reach magnitudes high enough to advance the contact line even in the absence of external agitation \cite{wasan_spreading_2003,nikolov_nanoparticle_2010,liu_dynamic_2012-1,liu_dynamic_2012,kondiparty_wetting_2011,chengara_spreading_2004}. This pressure scales with the number density of the nanoparticles meaning that higher volume fractions and smaller particles will drive the interface at faster rates. 

Despite the relative promise of SDP for oil removal there have been relatively few studies of its effectiveness. Submerging a core sample in a concentrated nanofluid, researchers found that a silica nanofluid increased oil recovery by a factor of 18 over DI water\cite{zhang_enhanced_2014,zhang_enhanced_2016}. Extending these experiments to a sintered bead pack and varying the capillary number, these researchers found a much smaller effect on the order of 21\% over brine solutions. By varying the capillary number of the nanofluid injection they found that high capillary number flows resulted in no improvement over an injected brine solution.  However, these experiments lacked the ability to visualize the interface motion, which is key to the SDP-driven oil removal. Further, most oil removal applications will drive the nanofluid with a bulk pressure gradient. Therefore, there is a need to examine how an external pressure gradient might influence the SDP-driven oil removal effectiveness of concentrated nanofluids.  \\

Here we present experiments that examine the ability of concentrated nanofluids to displace oil within a model microfluidic porous network while driven by a constant flow-rate pump. The oil/nanofluid interface is visible within the microfluidic chip, allowing observations of interface velocity and visualizations of the percent of oil displaced as a function of time. 
\section{Materials and Methods}\label{sec:intro}
\subsection{Model Microfluidic System}


\begin{figure}[h]
\centering

  \centering
  \includegraphics[width=0.95\textwidth]{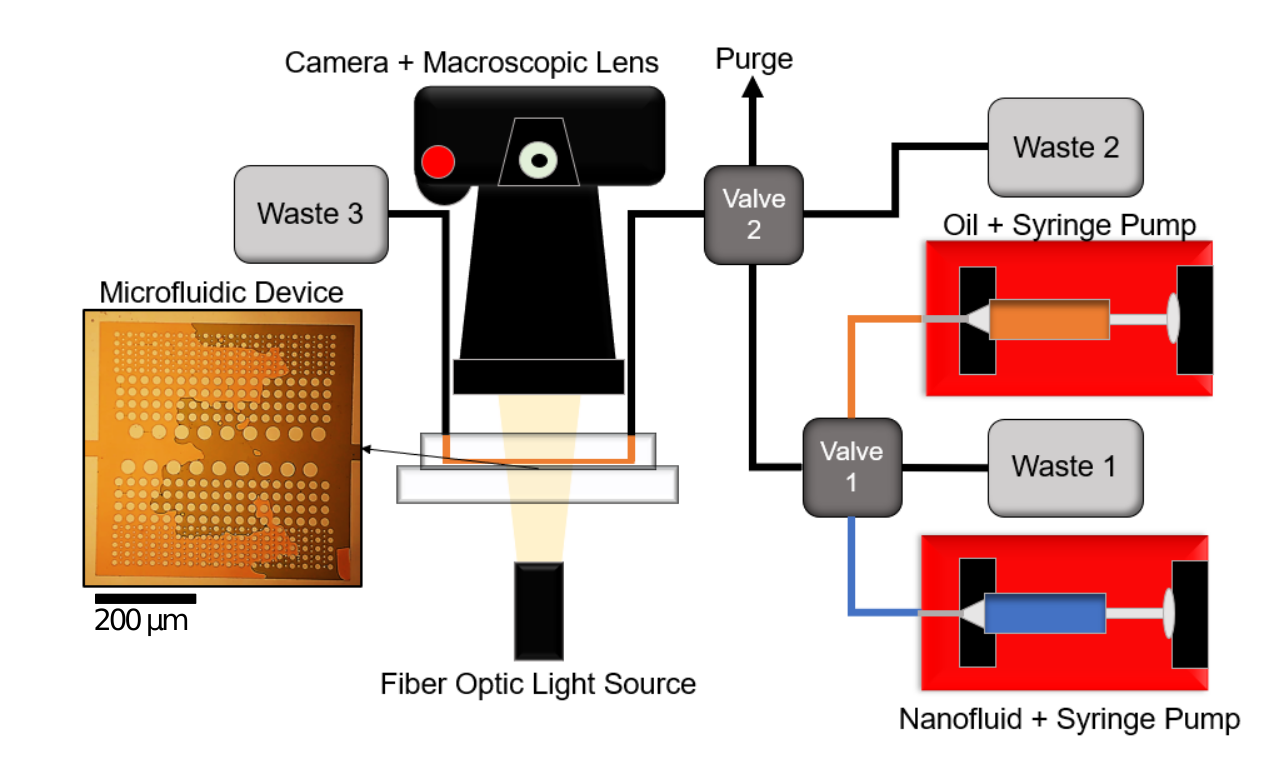}
\caption{Schematic showing the syringe pumps, valves, and arrangement of microfluidic porous chamber and imaging/illumination system.    }
\label{fig:experiment_schematic}
\end{figure}%

\begin{figure}
  \centering
  \includegraphics[width=1.0\textwidth]{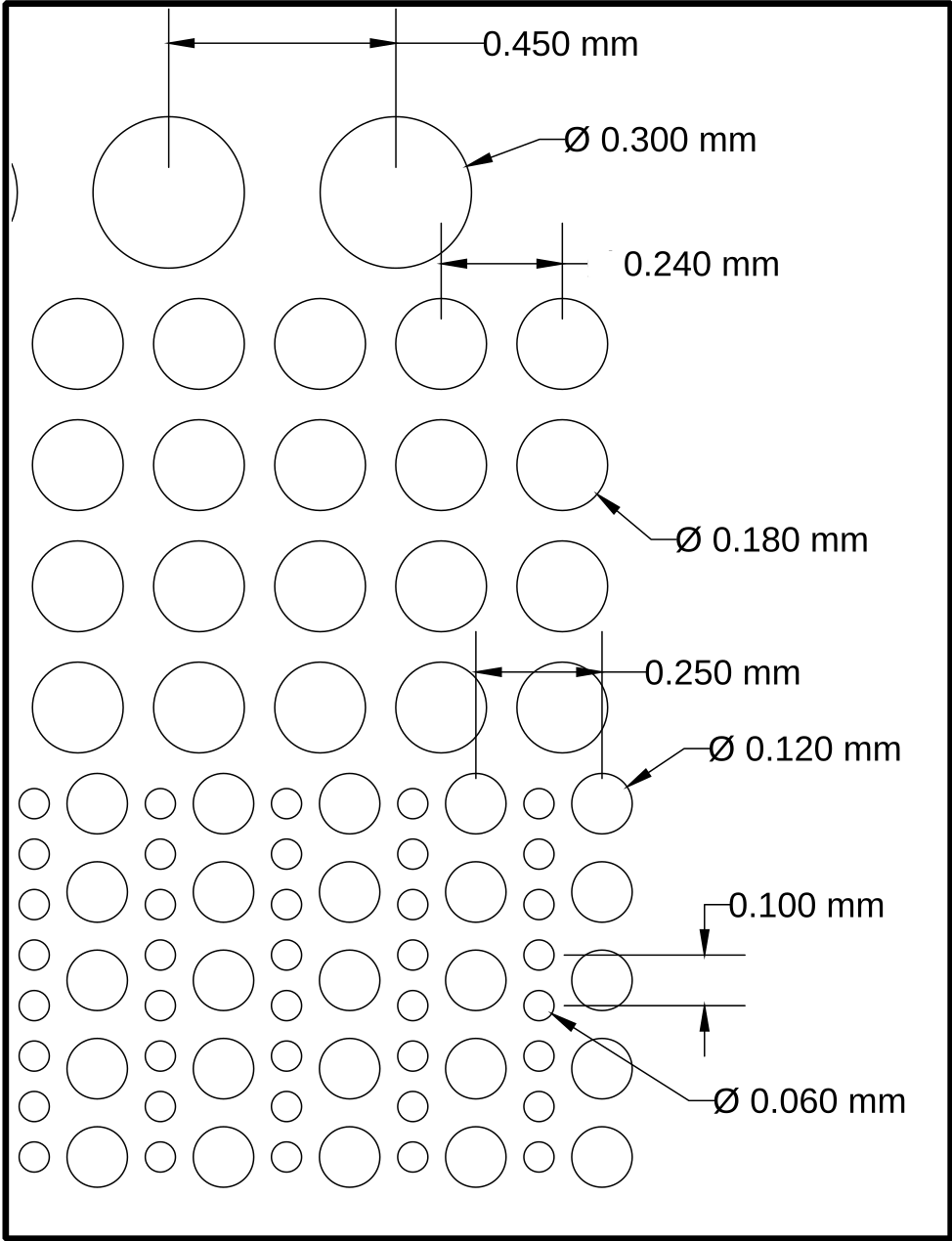}
\caption{Porous network pillar diameters and spacing, channel depth is $\mathrm{40~ \mu m}$.  }
\label{fig:dimensions}
\end{figure}

Figure \ref{fig:experiment_schematic} shows a schematic of our experimental system. Oil is flooded into the model microfluidic porous system using a syringe pump followed by the micelle nanofluid. Two valves are used to direct the flow and ensure that a clean interface is formed between the oil and nanofluid with no bubbles. Once the oil/nanofluid interface reaches the microfluidic chamber, its progress is imaged using a camera with a macroscopic lens, an example image of which is shown in the inset. The microfluidic channel then drains into a waste container at atmospheric pressure. Note that care is taken to ensure all components are at equal height to minimize flows induced by gravitational potential energy. 

The geometry of the microfluidic system was designed to mimic porous rock structure, which has been simulated as a porosity between .45 and .75 in similar oil recovery scenarios \cite{rosero_design_2018}. To achieve this goal, the system was designed to have inlets 300 $\mu$m wide and a square cavity 5 mm by 5 mm, which has a gridded circular column structure with a central channel of the same width as the inlets. This central channel was intended as a high permeability zone so that flow coming through the square cavity would be able to pass through the device without passing through the low permeability porous pillar structure. The porous pillar structure has pillars of decreasing size as one approaches the edge of the square cavity, ranging from 300 $\mu$m to 60 $\mu$m, with decreasing permeability towards the edge. These features result in a porosity of .75 for the microfluidic system as a whole. Rosero et al. utilized a very similar structure in testing polymer flooding EOR with a porosity of .61, albeit with octagonal pillars rather than circular ones \cite{rosero_design_2018}.


To create the geometry of our microfluidic system, SU-8 2025 Negative Photoresist (Kayaku Advanced Materials) was spin-coated to a height of  $\mathrm{40~ \mu m} $ using a program starting at a speed of 500 rpm for 15 s, then a speed of 2000 rpm for 1 minute to achieve a 40 $\mu$m height, and a rampdown to 500 rpm for another 15 s. The SU-8 was then exposed to $\mathrm{200~ mJ/cm^{2} }$ of UV light under a mask to create an inverse of the channel geometry which became our mold. The SU-8 mold was then silanized using a trichlorosilane treatment by allowing the trichlorosilane to evaporate in a covered container creating a self-assembled monolayer on the SU-8. This reduces the adhesion of PDMS to the SU-8 mold \cite{wang_microfabrication_2017}. 

Using this mold, the devices themselves were then fabricated out of polydimethylsiloxane (PDMS) using the SYLGARD 184 Silicone Elastomer Kit (Dow Chemicals) and following the manufacturer-provided fabrication procedure. Once the devices were bonded to a second flat sheet of PDMS, a hydrophilic treatment of polyvinyl alcohol was flown through the channel and then evaporated in the oven at $\mathrm{95^{o}}$C for 1 hour to render the channels hydrophilic \cite{park2021microfluidic,doan2023shape}. Then, in order to prevent the SDP from pushing the oil from the channel into the medium of the semi-permeable PDMS during the experiments, the devices were all soaked in canola oil for 2 days minimum in order to saturate the oil into the PDMS matrix.\\

\subsection{Nanofluid Preparation}

The micellar nanofluid used in the experiments was prepared using only Triton X-100 (Sigma-Aldrich) and DI water in different volume fractions. The following volume fractions were prepared: $\mathrm{ 2 \% , 4 \% , 10 \% , 15 \% , 20 \%,}$ and $\mathrm{30\%}$. These micellar solutions were mixed and dyed in different colors, then sonicated in an ultrasonic cleaner for 5 minutes to ensure proper formation of the micelles.

\begin{table}[h]
\begin{center}
\begin{minipage}{\textwidth}
\caption{Micelle hydrodynaimc diameter vs. volume fraction }\label{tab2}
\begin{tabular*}{\textwidth}{@{\extracolsep{\fill}} c c  c @{\extracolsep{\fill}}}
\toprule%

Volume fraction (\%) & Hydrodynamic diameter (nm) & Diameter standard deviation (nm) \\
\midrule
2 & 8.592 & 1.927 \\
\hline
4 & 8.076 & 1.746 \\
\hline
10 & 7.623 & 1.795 \\
\hline
15 & 7.390 & 1.809 \\
\hline
20 & 6.723 & 1.590 \\
\hline
30 & 5.209 & 1.234 \\
\hline
\label{tab:micelles}
\end{tabular*}
\end{minipage}
\end{center}
\end{table}

Each micellar nanofluid was characterized using dynamic light scattering (Litesizer 500, Anton-Parr) to verify their properties and find the average hydrodynamic diameter, the results of which are summarized in Table \ref{tab:micelles}. Triton X-100 micelle diameter is reported to be on the order of ~7.5-10 nm \cite{paradies_shape_1980}, and it was found that the average hydrodynamic diameter of the Triton X-100 micelles for each volume fraction was within the expected range. The relatively constant ratio of micelle diameters and standard deviations indicates a low polydispersity among the samples that were measured. This serves as an indication that experiments using these solutions have provided accurate and repeatable results. 

\subsection{Image Acquisition}

To acquire images, a Canon EOS R10 Mirrorless camera equipped with a Laowa 100 mm f/2.8 Macroscopic Lens was placed on a tripod above the microfluidic device with a white light fiber optic light source illuminating from below. The aperture of the lens was kept at f/2.8 so as to keep the depth of field as shallow as possible over the microfluidic device. The images were acquired on a time-lapse of 5 seconds with a total time of 5 hours as the experiment ran. This was found to be an appropriate time-lapse interval and total time for the activity seen in the experiment. 

\subsection{Image Analysis}

Images were analyzed procedurally using Python and OpenCV. To begin the analysis, the image is converted to grayscale and a 3-pixel width Gaussian blur is applied in order to remove high-frequency sensor noise. Oil displacement at a given frame is calculated on a per-pixel basis, with a pixel being considered displaced when the ratio between its current brightness and its brightness on the first frame exceeds 0.05 in the color channel corresponding to the oil dye color. 

There is often a significant delay between the start of the time-lapse and the time at which fluid actually enters the area of interest. The data are aligned by disregarding all frames with no significant change, such that the new first frame is the one in which the rate of area displaced exceeds 0.06\% of the channel's total area per second. 

To limit computation time, 
not every frame of the time-lapse is considered. The time--boundary plots (Figs. \ref{fig:oil_0} and \ref{fig:oil_30}) use every 24th frame, and the time--displacement plots (Fig. \ref{fig:percent_v_time_all})use every 5th frame. This correlates to time intervals of 120 seconds and 25 seconds, respectively.

In the generation of plots, portions of the frame outside the area of interest need to be disregarded. The corners of the channel are manually selected, and a mask is generated to not count pixels that are not within the bounds of the channel. The corner positions are later used in order to rotate and crop the data as shown in the time--boundary plot. 

\section{Results}\label{sec2}

\begin{figure}[h]

  \centering
  \includegraphics[width=1.0\linewidth]{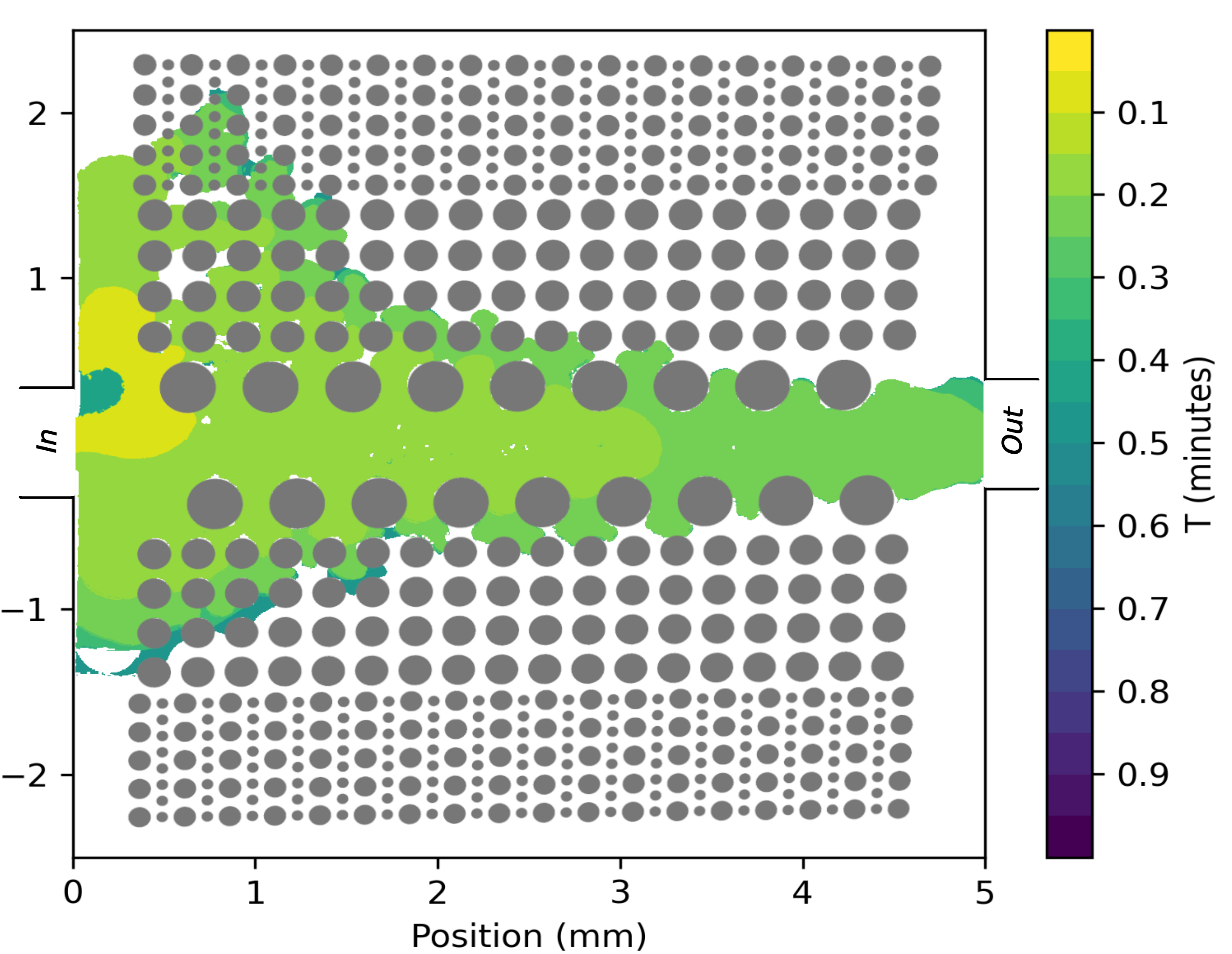}
  \caption{Time evolution of DI Water displacing oil injected at a $\mathrm{10~ \mu L/hr.}$ flow rate. }
  \label{fig:oil_0}
\end{figure}%
\begin{figure}
  \centering
  \includegraphics[width=1.0\linewidth]{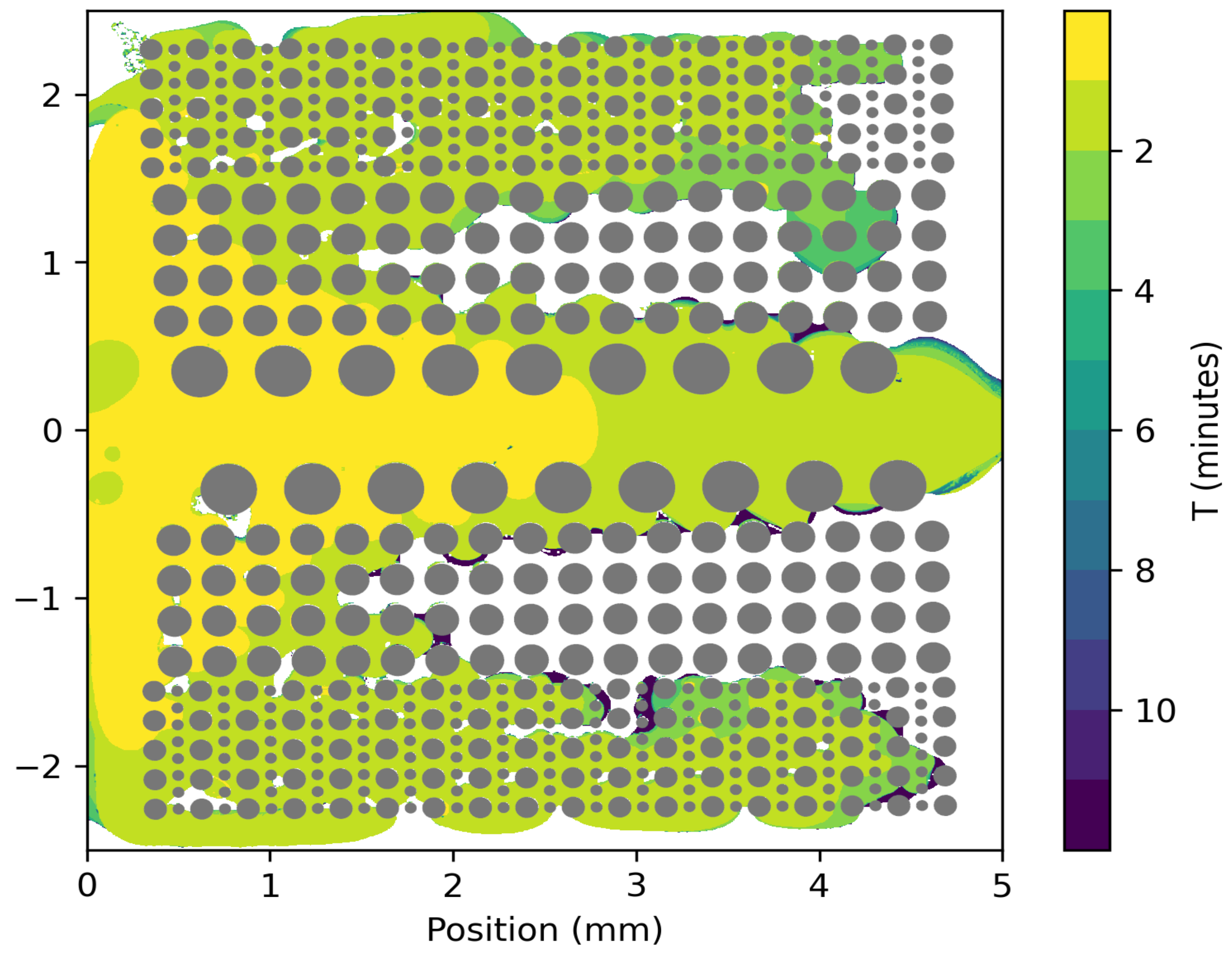}
  \caption{Time evolution of a 30\% Volume Fraction Nanofluid displacing oil in a porous channel, injected at a flow rate of $\mathrm{10~ \mu L/hr.}$.}
  \label{fig:oil_30}
\end{figure}

This experimental setup allows for easy visualization of the interface progress, as can be seen in Figures \ref{fig:oil_0} and \ref{fig:oil_30}. Here we plot the location of the nanofluid/oil interface as a function of time for both deionized (DI) water and a 30\% volume fraction nanofluid. The entrance and exit to the model porous media section are labeled and the pillar structure of the microfluidic channel is overlaid to facilitate interpretation of the results. 

The results in Figures \ref{fig:oil_0} and \ref{fig:oil_30} show that the concentrated nanofluid is both effective at displacing the oil and actually more effective in the less permeable regions of the channel. In Figure \ref{fig:oil_30}, one can see the nanofluid enter from the left and then spread to form three advancing fronts: one in the middle, high porosity section, and two in the least porous sections at the top and bottom. The middle front corresponds to the traditional, pressure gradient-driven displacement modality, while the upper and lower branches are SDP driven. The SDP-driven interfaces in this case were almost as fast as the pressure gradient-driven flow, largely because the less porous sections have far more three-phase contact line length. Each contact line contributes a force/length due to the SDP and the increased opportunities to form contact line results in more force per interfacial area driving the flow forward.

\begin{figure}[h]%
\centering
\includegraphics[width=0.7\textwidth]{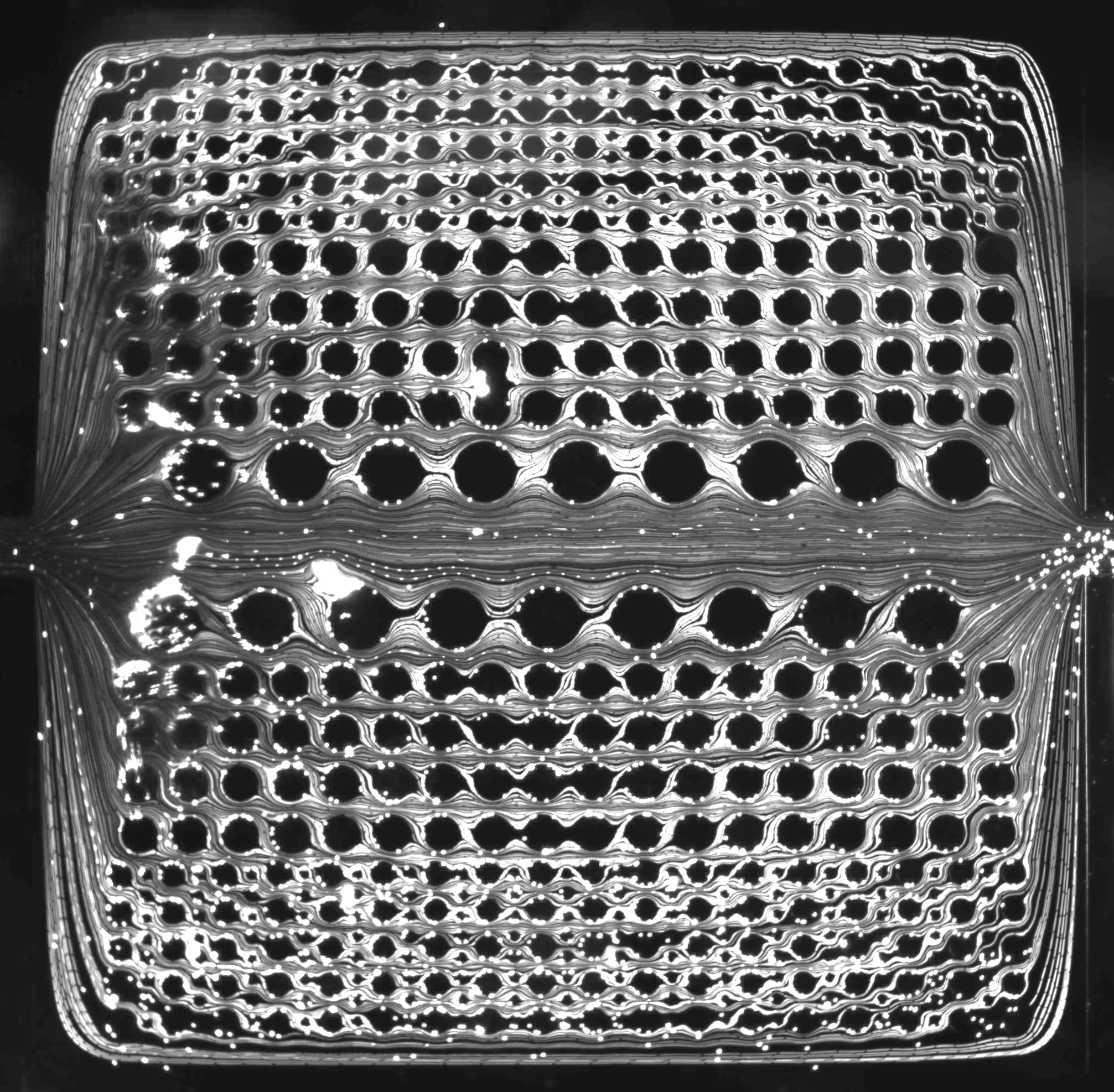}
\caption{Plot of fluorescent microsphere streaklines showing streamlines for water flowing through the channel with a flow rate of $\mathrm{10~ \mu L/hr}$. As expected from the low Reynolds number flow, streaklines follow the symmetry of the model microfluidic porous network.  }
\label{fig:flow_vis}
\end{figure}

Figure \ref{fig:flow_vis} further supports that the nanofluid is more effective at displacing oil in the less porous sections of the channel. Here we show 50 streakline plots of $\mathrm{7.32 ~ \mu m}$ fluorescent polystyrene spheres flowing in our model microfluidic porous media, combined into a single image using the \textit{ImageJ} plugin \textit{Flowtrace}. Comparing this figure with the interface trajectory displayed in Figure \ref{fig:oil_30}, one can see that indeed the SDP-driven interface does not follow the flow path and instead moves quickly into the low porosity region and stays there. 

In contrast to the concentrated nanofluid, the water in Figure \ref{fig:oil_0} behaves largely as expected. The vast majority of the oil displacement occurs in the middle, high porosity section of the microfluidic channel. The water does displace a small amount of oil from the medium porosity sections of the channel but this is likely a result of the shallow height of the channel. The $\mathrm{40~ \mu m}$ depth of the channel is smaller than the $\mathrm{60~ \mu m}$ pillar spacing in this section of the channel making the pillar's contribution to the flow resistance negligible. It isn't until the outer, low porosity section that the pillar spacing becomes on the order of the channel depth, making their contribution to the viscous resistance significant. 

\begin{figure}[h]%
\centering
\includegraphics[width=0.9\textwidth]{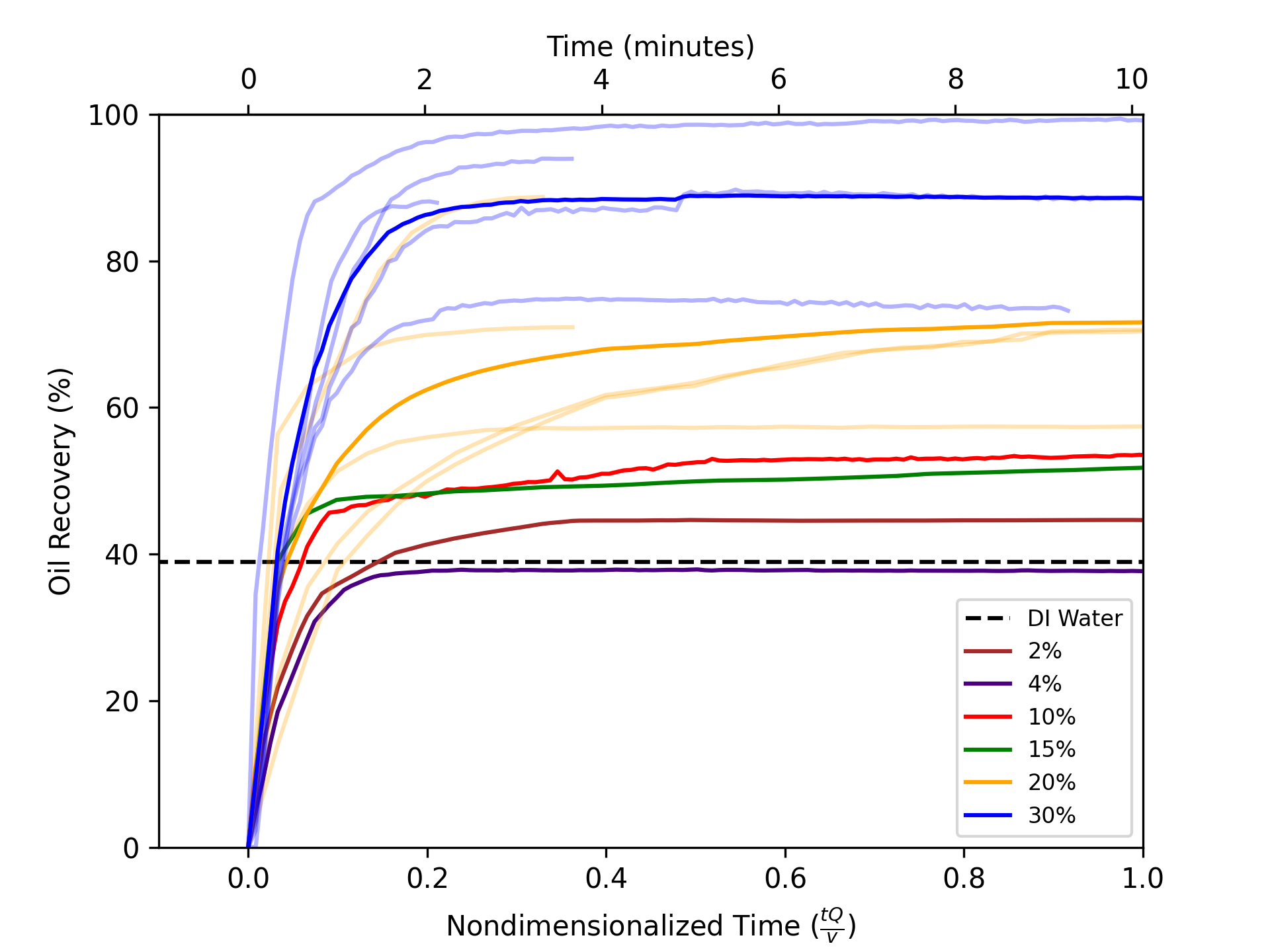}
\caption{Plot of \% oil displaced as a function of time for a range of different nanofluid volume fractions. For higher volume fractions, 5 trials are shown (light lines) along with their average (dark lines). For lower volume fractions only the average is shown for readability as they highly overlap. }
\label{fig:percent_v_time_all}
\end{figure}

As the volume fraction of the nanofluid increases, so does the percentage of oil displaced as shown in Figure \ref{fig:percent_v_time_all}. Here five experimental runs are shown as light lines with the average plotted as a darker version for nanofluid volume fractions of 20\% and 30\%. For lower volume fractions only the average is plotted. Note that the dimensionless time, normalized by the flow rate divided by the model porous media chamber volume is provided alongside the real-time (min). The individual experimental runs are shown to highlight the variability of the results: the 30\% volume fraction nanofluid cleared anywhere from 73\%  to 99\% of the oil for example while the 20\% nanofluid varied from 55\% to 86\% oil cleared. This variability in oil displacement is likely a result of heterogeneity in the channel surface energy, contact line pinning seems to strongly affect the route the nanofluid takes and hence its speed and the amount of oil removed. 

Another notable feature of this plot is how a high percentage of oil displaced only occurs when the SDP-driven displacement is completed in a dimensionless time on the order of 1. For volume fractions of 0\% through 15\% the percentage of oil removed can be seen to slowly increase due to SDP after an initial rapid rise due to the pressure gradient-driven displacement. This rise occurs over dimensionless time that extends beyond the experimental window, i.e. much higher than 1.  For the 20\% volume fraction the work of the nanofluid is completed on the order of a single dimensionless time. Indeed, from the individual experiment lines it is clear that the initial pressure gradient and the SDP-driven displacement are seamless. In other words, the SDP-driven interface velocity must be on the order of the flow-driven velocity for SDP to significantly enhance the oil displacement in a model porous network.

\begin{figure}[h]%
\centering
\includegraphics[width=0.9\textwidth]{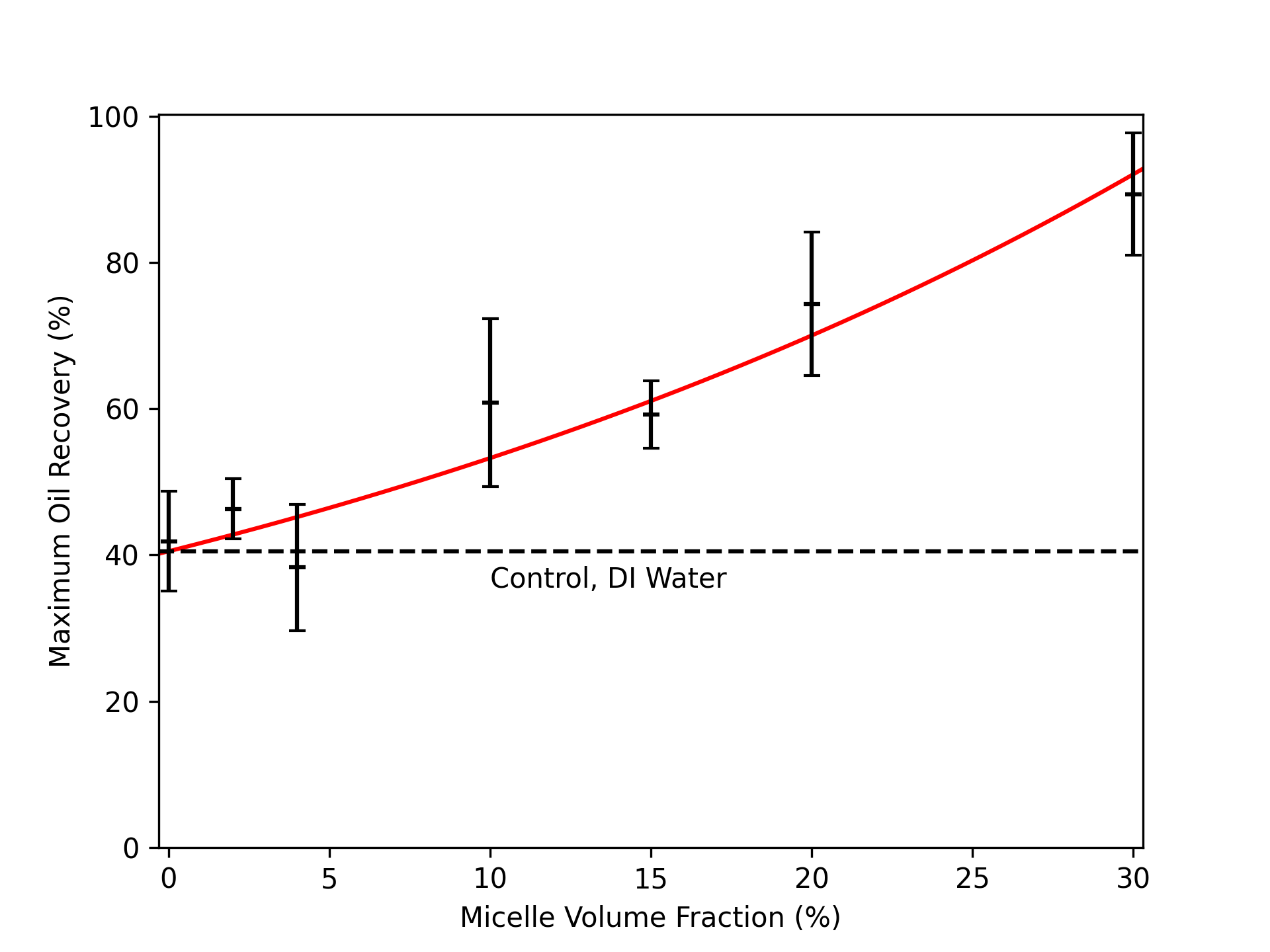}
\caption{Plot of final \% oil displaced as a function of nanofluid volume fraction. Solid red line is an exponential fit ($OR=e^{0.0274 VF + 3.701}, ~~ \mathrm{R^2=0.9602} $ )}
\label{fig:percent_v_VF}
\end{figure}

Regardless, increasing the volume fraction of micelles increases the maximum of oil removed as can be seen in Figure \ref{fig:percent_v_VF}. Here we show the maximum percentage of oil displaced as a function of nanofluid volume fraction and attempted to fit an exponential curve to the data, the equation of which is given in Figure \ref{fig:percent_v_VF} caption where $VF$ is the nanofluid volume fraction and $OR$ is the oil recovery percentage. The result is an almost linear increase in oil displaced with increasing volume fraction, though this behavior should not be taken as broadly applicable. While we did not examine the effect of model porous media geometry it is almost certain to affect these results. For example, a smaller center, high porosity section combined with a deeper channel would likely result in a lower amount of oil recovered by DI water by exaggerating both the change in resistance between different sections of the channel and lowering the volume of the low resistance flow path. The end result would be to make the nanofluid appear more effective at displacing oil. 

\section{Conclusion}
We have utilized a model, porous media microfluidic channel to assess the oil-removing capabilities of the SDP while pumping the fluid at a constant flow rate. The oil displacing capabilities of the nanofluid improve with increasing volume fraction as would be expected from prior work. This improvement is roughly linear with nanofluid volume fraction, which is also in line with prior work which has shown that the interface velocity for a sessile oil drop is also modeled as linear with nanoparticle volume fraction\cite{liu_dynamic_2012}, though to our knowledge this is the first validation of this modeling across such a wide range of volume fractions. What was unexpected in our results was the wide range of observed oil displacement percentages for a given nanofluid, likely showing the importance of surface energy heterogeneity in the SDP-driven oil displacement process.

\section{Conflicts of Interest}
The authors declare no conflicts of interest. 

\section{Funding Declaration}
The authors would like to acknowledge the University at Buffalo for supporting this work.


\bibliographystyle{tfnlm}
\bibliography{bibliography}

\end{document}